\documentclass[%
 aip,
 numerical,
 jcp,
rsi,%
amsmath,amssymb,
preprint,%
author-year,%
]{revtex4-1}
\usepackage{xcolor}
\usepackage{graphicx}% Include figure files
\usepackage{dcolumn}% Align table columns on decimal point
\usepackage{bm}
\usepackage{natbib}
\usepackage{hyperref}
\usepackage{amsmath}
\usepackage{amssymb} 
\usepackage{braket}
\usepackage{mathtools}
\usepackage{nicefrac}
\usepackage{bm}

\newcommand{\diag}{\mathop{\mathrm{diag}}}\bibpunct{(}{)}{,}{n}{,}{,}

\newcommand{\sgn}{\text{sgn}}

\begin{document}

%\usepackage[mathlines]{lineno}% Enable numbering of text and display math
%\linenumbers\relax % Commence numbering lines
\title{The use of the Born-Oppenheimer factorization in the phase-space representation of the time-independent Schrödinger equation for bilinearly coupled harmonic oscillators.}

\author{Carlos A. Arango}
\email{caarango@icesi.edu.co}
\affiliation{Department of Chemical Sciences, Universidad Icesi, Cali, Colombia}

\date{\today}

\begin{abstract}

A system of two bilinearly coupled harmonic oscillators has been solved analytically by using the Born-Oppenheimer (BO) product wavefunction ansatz and the phase-space bound trajectory approach [J. S. Molano et al., Chem. Phys. Lett. \textbf{76}(12), 138171 (2021)]. The bilinearly coupled oscillator system allows to obtain the analytical expression of the quantum system, facilitating comparison with the results of using the BO ansatz product. The analytical and BO wavefunctions are obtained as a product of parabolic cylinder functions. The arguments of the parabolic cylinder functions of the exact and the BO wavefunctions are related to the physical coordinates by linear transformations, represented by matrices $\mathsf{G}$ and $\mathsf{\tilde{G}}$ respectively. A $QU$ decomposition of the
matrix $\mathsf{G}$ outputs an upper triangular matrix $\mathsf{U}$ that is closely related to the BO $\mathsf{\tilde{G}}$ matrix. The effect of the BO non-adiabatic coupling is analyzed on the stability of the phase-space equations of motion. The eigenvalues and eigenfunctions obtained by the Born-Oppenheimer phase-space trajectory approach show excellent agreement with the analytical solutions. 

\end{abstract}


\maketitle 

\section{Introduction}

Coupled oscillators models have been extensively used in chemical physics \cite{Uzer1991,Rice2009}. Intramolecular vibrational energy transfer \cite{Uzer1991,Rice2009}, vibrational spectroscopy of polyatomic molecules \cite{Herzberg1990,sathyanarayana2015}, and quantum entanglement \cite{Makarov2018,Chakraborty2018} are among the successful applications in chemical physics. Bilinearly coupled harmonic oscillators have been vastly studied and used to represent physical systems due to the possibility of having analytical solutions of the Schr\"odinger equation by using algebraic decoupling approaches and generalized transformations among other mathematical techniques \cite{Moshinsky1996,Feldman2000,Burrows2000,Burrows2003,Zuniga2017}. Approximate solution methods based on the self consistent field (SCF) theory and the adiabatic approximation (AA) have been used to solve the Schr\"odinger equation for coupled harmonic oscillators \cite{Caswell1970,Bowman1978,Qian1982,Ezra1983,Certain1987,Makarewicz1985,Craig1994}. These method are based on product factorizations of the wavefunction. Self consistent field is based on Hartree products, meanwhile the adiabatic approximation method is based on Born-Oppenheimer products. The SCF is corrected by allowing state interaction \cite{Bowman1979}. The AA is improved by including the non-adiabatic corrections by means of perturbation theory \cite{Born1927,Frey1988,Craig1994}.

In this work a phase-space trajectory method, proposed previously \cite{Molano2021}, for solving the time independent Schrödinger equation (TISE) has been extended to systems of two degrees of freedom. This phase-space method is based on the use of bound trajectories in phase-space as eigenstates of the TISE. The extension of the phase-space method to systems with two degrees of freedom has been done by using the Born-Oppenheimer product ansatz for the wave function. The equations developed by the use of the phase-space trajectory approach are applied to a system of two harmonic oscillators with bilinear coupling. The resulting differential equations can be written in matrix form allowing to study the dynamical structure and stability of the system \cite{Strogatz1994}. Exact and Born-Oppenheimer wavefunctions are written in terms of parabolic cylinder functions $D_\nu$ \cite{abramowitz1964}. The arguments of the $D_\nu$ functions for the exact and BO solutions are given as linear combination of the oscillators coordinates, with well defined transformation matrices. The analysis of the elements of these transformation matrices allows to relate the error in the BO approximation with the coupling between the oscillators.  

\section{Theory and Methods}

A system of two interacting particles, labeled 1 and 2, with coordinates $x$ and $y$, respectively, is described by the time independent Schrödinger equation (TISE) $\hat{H}\psi(x,y)=\mathcal{E}\psi(x,y)$.
The Hamiltonian for this system can be written as
\begin{equation}\label{eq:full_hamiltonian}
   \hat{H}=\hat{T}_1(\hat{p}_x)+\hat{T}_2(\hat{p}_y)+\hat{V}_{12},
\end{equation}
with kinetic energy operators $\hat{T}_1=-\tfrac{1}{2m_1}\partial^2_x$, and $\hat{T}_2=-\tfrac{1}{2m_2}\partial^2_y$, and potential energy operator $\hat{V}_{12}=\hat{V}_{12}(x,y)$. This Hamiltonian can be separated as $\hat{H}=\hat{T}_2+\hat{h}$, with $\hat{h}=\hat{h}(\hat{p}_x,x,y)$ given by
\begin{equation}\label{eq:small_hamiltonian}
    \hat{h}=\hat{T}_1+\hat{V}_{12}.
\end{equation}
The Born-Oppenheimer approximation resorts to a product wave function, $\psi=\chi\varphi$, of the form 
\begin{equation}\label{eq:product_wavefunction}
    \psi(x,y)=\chi(x,y)\varphi(y).
\end{equation}
The $\chi(x,y)$ functions are obtained as the bound solutions of the $y$-parametrized TISE
\begin{equation}\label{eq:TISE_h}
\hat{h}\chi_n(x,y)=\varepsilon_n(y)\chi_n(x,y).
\end{equation}
The eigenvalues $\varepsilon_n(y)$ are the potential energy curves of the $y$ degree of freedom. The eigenfunctions $\chi_n(x,y)$ obey the orthonormality condition,
\begin{equation}\label{eq:orthonormality_chi}
    \braket{\chi_l|\chi_n}=\int_{-\infty}^{\infty}{\chi_l^*(x,y)\chi_n(x,y)dx}=\delta_{ln}(y),
\end{equation}
with $\delta_{ln}(y)$ as the $y$ dependent Kronecker delta.  

The action of the Hamiltonian operator on the wave function $\psi$, $\hat{H}\psi=\mathcal{E}\psi$, produces
\begin{equation}\label{eq:TISE}
    -\tfrac{1}{2m_1}\varphi\partial^2_x\chi-\tfrac{1}{2m_2}\left(\varphi\partial^2_y\chi+\chi\varphi''+2\varphi'\partial_y\chi\right)+\hat{V}\chi\varphi=\mathcal{E}\chi\varphi,
\end{equation}
with $\mathcal{E}$ as the total energy of the full system. This result can be rearranged to,
\begin{equation}\label{eq:TISE_rearranged}
-\tfrac{1}{2 m_2}\chi\varphi''=\left(\tfrac{1}{2m_1}\partial^2_x\chi-(\hat{V}-\mathcal{E})\chi\right)\varphi+\tfrac{1}{2m_2}\left(2\varphi'\partial_y\chi+\varphi\partial^2_y\chi\right).
\end{equation}
The last term of this equation can be written compactly as 
\begin{equation}\label{eq:derivative_rearrangement}
    \frac{1}{2m_2}\left(2\varphi'\partial_y\chi+\varphi\partial^2_y\chi\right)=\frac{\partial_y(\varphi^2\partial_y\chi)}{2m_2\varphi},
\end{equation}
which can be used, together with the definition of $\hat{h}$, in equation \eqref{eq:TISE_rearranged} to get
\begin{equation}\label{eq:TISE_using_h_and_derivative}
   -\tfrac{1}{2 m_2}\chi\varphi''=(\mathcal{E}-\hat{h})\chi\varphi+\frac{\partial_y(\varphi^2\partial_y\chi)}{2m_2\varphi}.
\end{equation}
The use of the eigenfunctions $\chi_n$ in equation \eqref{eq:TISE_using_h_and_derivative} gives
\begin{equation}\label{eq:TISE_with_chi_n}
  -\tfrac{1}{2 m_2}\chi_n\varphi_n''=\left(\mathcal{E}-\varepsilon_n(y)\right)\chi_n\varphi_n+\frac{\partial_y(\varphi_n^2\partial_y\chi_n)}{2m_2\varphi_n}.
\end{equation}
A simpler differential equation for $\varphi_n(y)$ is obtained if the last term on the right of \eqref{eq:TISE_with_chi_n} is zero, which is possible only if $\varphi_n^2\partial_y\chi_n=f_n(x)$, with $f_n(x)$ as a function of $x$ only. Integrating this last result produces
\begin{equation}\label{eq:approx_y_ode}
    \chi_n=g_n(x)+f_n(x)\int{\frac{dy}{\varphi_n^2}}.
\end{equation}
with $g(x)$ as an arbitrary function of $x$. The result \eqref{eq:approx_y_ode} indicates that the second term of equation \eqref{eq:TISE_with_chi_n} can be left out in the case of complete separation of variables.

The use of the definition $\vartheta_n=\varphi'_n$, allows to write equation \eqref{eq:TISE_with_chi_n} as a system of equations, for given $x$ and $n$,
\begin{equation}\label{eq:var_equations}
    \begin{split}
    \varphi_n'&=\vartheta_n,\\
    \chi_n\vartheta_n'&=\left(-2m_2\left(\mathcal{E}-\varepsilon_n(y)\right)\chi_n-\partial^2_y\chi_n\right)\varphi_n-(2\partial_y\chi_n)\vartheta_n.
    \end{split}
\end{equation}
These are the equations for the phase-space of $\varphi_n$ for a given value of the coordinate $x$.

Multiplying the second equation \eqref{eq:var_equations} on the left by $\chi_l^*$ and integrating over $x$ gives
\begin{equation}\label{eq:vartheta_equation}
    \delta_{ln}\vartheta_n'=\left(-2m_2(\mathcal{E}-\varepsilon_n(y))\delta_{ln}-\beta_{nl}\right)\varphi_n-2\alpha_{nl}\vartheta_n,
\end{equation}
with $\alpha_{nl}=\braket{\chi_l|\partial_y\chi_n}$ and $\beta_{nl}=\braket{\chi_l|\partial_y^2\chi_n}$ as the first and second order non-adiabatic couplings, respectively.  

The diagonal part of equation \eqref{eq:vartheta_equation} gives
\begin{equation}\label{eq:vartheta_equation_diagonal}
    \vartheta_n'=\left(-2m_2(\mathcal{E}-\varepsilon_n(y))-\beta_{nn}\right)\varphi_n-2\alpha_{nn}\vartheta_n,
\end{equation}
with $\alpha_{nn}=\braket{\chi_n|\partial_y\chi_n}$ as the Berry's connection. Since the eigenfunctions $\chi_n$ are real and normalized, we have $\alpha_{nn}=0$, and
\begin{equation}\label{eq:vartheta_equation_diagonal_2}
    \vartheta_n'=\left(-2m_2(\mathcal{E}-\varepsilon_n(y))-\beta_{nn}\right)\varphi_n.
\end{equation}

The off-diagonal elements of equation \eqref{eq:vartheta_equation} give the equations
\begin{equation}\label{eq:vartheta_equation_off_diagonal}
    -\beta_{nl}\varphi_n=2\alpha_{nl}\vartheta_n.
\end{equation}

By defining the vector $\mathbf{\Phi}_n=\{\varphi_n,\vartheta_n\}$, the diagonal equation \eqref{eq:vartheta_equation_diagonal} can be written in matrix form $\mathbf{\Phi}'_n=\mathsf{A}_n\mathbf{\Phi}_n$ with 
\begin{equation}\label{eq:matrix_A}
    \mathsf{A}_n=\begin{pmatrix}
    0 & 1 \\
    -2m_2\left(\mathcal{E}-\varepsilon_n(y)\right)-\beta_{nn}
     & -2\alpha_{nn}\\
     \end{pmatrix}.
\end{equation}
The system of equations $\mathbf{\Phi}'_n=\mathsf{A}_n\mathbf{\Phi}_n$ displays a fixed point at $\mathbf{\Phi}_n=0$, with stability given by the eigenvalues of matrix $\mathsf{A}_n$,
\begin{equation}\label{eq:eigenvalues_A}
    \lambda_{1,2}=-\alpha_{nn}\mp\sqrt{\alpha_{nn}^2-2m_2\left(\mathcal{E}-\varepsilon_n(y)\right)-\beta_{nn}}.
\end{equation}
Elliptic phase-space trajectories around the fixed point are given by values of $y$ where the condition $\alpha_{nn}^2-2m_2\left(\mathcal{E}-\varepsilon_n(y)\right)-\beta_{nn}<0$ is fulfilled. Unlike 1-dimensional systems, classical turning points are not boundaries between elliptic and hyperbolic behaviour of the phase-space trajectories \cite{Molano2021}. Instead, 2-dimensional systems display non-adiabatic turning points given by the $y$ solutions of $\alpha_{nn}^2-2m_2\left(\mathcal{E}-\varepsilon_n(y)\right)-\beta_{nn}=0$.

\section{Bilinearly coupled 1-dimensional harmonic oscillators}

Consider two 1-dimensional harmonic oscillators of frequencies $\omega$ and $\Omega$, described by coordinates $x$ and $y$, and with masses $m$ and $M$, respectively. The oscillators are coupled by a bilinear term, $cxy$. The Hamiltonian operator is given by
\begin{equation}\label{eq:Hamiltonian_xy}
\hat{H}=-\frac{1}{2m}\frac{\partial^2}{\partial x^2}-\frac{1}{2M}\frac{\partial^2}{\partial y^2}+\frac{m\omega^2}{2} x^2+\frac{M\Omega^2}{2} y^2+ c x y,
\end{equation}
It is convenient to describe the coupling by $\delta=c\left(\sqrt{mM}\omega\Omega\right)^{-1}$, restricted to $\left|\delta\right|<1$. 
The uncoupled case, $\delta=0$, gives as solution
\begin{equation}\label{eq:uncoupled_case}
    \Psi(x,y)=\frac{\left(mM\omega\Omega\right)^{1/4}}{\sqrt{\pi n_1!n_2!}}D_{n_1}(\sqrt{2m\omega}x)D_{n_2}(\sqrt{2M\Omega}y),
\end{equation}
where $D_\nu(\xi)$ are the parabolic cylinder (Weber) functions \cite{abramowitz1964}, and $\nu=n_1,n_2$, as non negative integers. 

\subsection{Analytical solution}

It is convenient to use the matrix notation for the Hamiltonian \eqref{eq:Hamiltonian_xy},
\begin{equation}\label{eq:Hamiltonian_xy_matrix}
    2\hat{H}=\mathbf{z}^\mathsf{t}\mathsf{H}\mathbf{z},
\end{equation}
with phase-space operator $\mathbf{z}$ as the column vector obtained by the concatenation of $\mathbf{r}=\begin{psmallmatrix}x \\ y\end{psmallmatrix}$ and $\nabla_\mathbf{r}=\begin{psmallmatrix}\partial_x \\ \partial_y\end{psmallmatrix}$. The matrix $\mathsf{H}$ is given by 
\begin{equation}\label{eq:H_matrix}
    \mathsf{H}=\begin{pmatrix} m\omega^2 & c & 0 & 0 \\
    c & M\Omega^2 & 0 & 0 \\
    0 & 0 & -\frac{1}{m} & 0 \\
    0 & 0 & 0 & -\frac{1}{M}\end{pmatrix}.
\end{equation}
The use of mass-weighted coordinates $\bm{\rho}=\sqrt{\mathsf{M}}\mathbf{r}$, with $\mathsf{M}=\diag(m,M)$, and the corresponding gradient operator $\nabla_{\bm{\rho}}=\frac{1}{\sqrt{\mathsf{M}}}\nabla_{\mathbf{r}}$, allows to define the mass-weighted phase-space operator $\bm{\zeta}$ as the concatenation of $\bm{\rho}$ and $\nabla_{\bm{\rho}}$. The use of $\bm{\zeta}$ simplifies equation \eqref{eq:Hamiltonian_xy_matrix} to
\begin{equation}\label{eq:Bar_Hamiltonian_xy_matrix}
    2\hat{H}=\bm{\zeta}^\mathsf{t}\mathsf{K}\bm{\zeta},
\end{equation}
with $\mathsf{K}=\diag\left(\mathsf{B},-\mathsf{I}\right)$ as a block diagonal matrix, with 
\begin{equation}\label{eq:Bar_H_matrix}
    \mathsf{B}=\begin{pmatrix} \omega^2 & \delta\omega\Omega \\
    \delta\omega\Omega & \Omega^2  \end{pmatrix},
\end{equation}
and $\mathsf{I}$ as the $2\times2$ identity matrix. 

Matrix $\mathsf{B}$ deals only with the potential energy part of the Hamiltonian \eqref{eq:Hamiltonian_xy}, $2V=\bm{\rho}^\mathsf{t}\mathsf{B}\bm{\rho}$.
The eigenvectors of $\mathsf{B}$ are rows of an orthogonal matrix $\mathsf{R}$ that diagonalizes $\mathsf{B}$. The transformation of coordinates $\mathbf{q}=\mathsf{R}\bm{\rho}$, with $\mathbf{q}=\begin{psmallmatrix}q_1 \\q_2\end{psmallmatrix}$, produces a diagonal matrix $\mathsf{L}$ similar to $\mathsf{B}$, $\mathsf{L}=\mathsf{RBR^{\mathsf{t}}}$, with $\mathsf{L}=\diag{\left(\lambda_1,\lambda_2\right)}$ as the matrix of eigenvalues of $\mathsf{B}$. The potential energy after this  transformation is given by $2V=\mathbf{q^t}\mathsf{L}\mathbf{q}$.

The eigenvalues of $\mathsf{B}$ are explicitly given by
\begin{equation}\label{eq:Bar_H_eval}
    \lambda_{1,2}=\tfrac{1}{2}\left(\omega^2+\Omega^2\pm\sqrt{\left(\omega^2-\Omega^2\right)^2+4\delta^2\omega^2\Omega^2}\right),
\end{equation}
which gives $0<\lambda_2<\lambda_1$ for $\Omega<\omega$. The normalized eigenvectors of $\mathsf{B}$ are the rows of the orthogonal matrix
\begin{equation}\label{eq:q_transformation}
\mathsf{R}
=\begin{pmatrix}
\cos{\theta} & \sin{\theta}\\
-\sin{\theta} & \cos{\theta}
\end{pmatrix},
\end{equation}
with $\theta=\arctan{\frac{\delta\omega\Omega}{\lambda_1-\Omega^2}}$, restricted to $\theta\in\left(-\nicefrac{\pi}{2},\nicefrac{\pi}{2}\right)$. The kinetic energy part of the Hamiltonian \eqref{eq:Hamiltonian_xy} requires the relationship between the gradient operators, $\nabla_{\mathbf{q}}=\mathsf{R}^{\mathsf{t}}\nabla_{\bm{\rho}}$. Since the transformation matrix $\mathsf{R}$ is orthogonal, the kinetic energy operator is finally given by $2\hat{T}=-\nabla_\mathbf{q}\cdot\nabla_\mathbf{q}$, and he Hamiltonian simplifies to
\begin{align}
    \hat{H}&=-\tfrac{1}{2}\nabla_\mathbf{q}\cdot\nabla_\mathbf{q}+\tfrac{1}{2}\mathbf{q^t}\mathsf{L}\mathbf{q}\\
    &=-\tfrac{1}{2}\left(\tfrac{\partial^2}{\partial{q_1^2}}+\tfrac{\partial^2}{\partial{q_2^2}}\right)+\tfrac{1}{2}\left(\omega_1^2q_1^2+\omega_2^2q_2^2\right),
\end{align}
with $\omega_{1,2}=\sqrt{\lambda_{1,2}}$. The transformation of coordinates from the original $\mathbf{r}$ to the $\mathbf{q}$ is given by $\mathbf{q}=\mathsf{R}\sqrt{\mathsf{M}}\mathbf{r}$ with volume element transforming as $dxdy=\frac{dq_1 dq_2}{\sqrt{m M}}$.

The analytical energy eigenvalues are:
\begin{equation}\label{eq:analytical_energy}
    \mathcal{E}_{n_1,n_2}=\omega_1\left(n_1+\tfrac{1}{2}\right)+\omega_2\left(n_2+\tfrac{1}{2}\right),
\end{equation}
with $n_{1,2}$  non negative integers. A simple inspection of equations \eqref{eq:Bar_H_eval} and \eqref{eq:analytical_energy}, for $\omega>\Omega$, shows that for the uncoupled case $\delta=0$, 
\begin{equation}\label{eq:analytical_energy_uncoupled}
    \mathcal{E}_{n_1,n_2}=\omega\left(n_1+\tfrac{1}{2}\right)+\Omega\left(n_2+\tfrac{1}{2}\right).
\end{equation}

The normalized analytical wave function is given by $\Psi_{n_1,n_2}(q_1,q_2)=\psi_{n_1}(q_1)\psi_{n_2}(q_2)$, with
\begin{equation}\label{eq:psi_analytical_wf}
    \psi_{n_i}(q_i)=\frac{1}{\sqrt{n_i!}}\left(\frac{mM\omega_i}{\pi}\right)^{1/4}D_{n_i}(\sqrt{2\omega_i}q_i),
\end{equation}
where $i=1,2$, and $D_{n_i}$ are the parabolic cylinder functions. The normalized wave function is given by
\begin{equation}
\begin{split}\label{eq:Psi_analytical_wf}
    \Psi(q_1,q_2)&=\frac{1}{\sqrt{n_1! n_2!}}\left(\frac{\omega_1\omega_2}{\pi^2}\right)^{1/4}D_{n_1}(\sqrt{2\omega_1}q_1)D_{n_2}(\sqrt{2\omega_2}q_2)\\
    &=\frac{\left(\omega\Omega\sqrt{1-\delta^2}\right)^{1/4}}{\sqrt{\pi n_1! n_2!}}D_{n_1}(\sqrt{2\omega_1}q_1)D_{n_2}(\sqrt{2\omega_2}q_2).
    \end{split}
\end{equation}
The arguments of the $D_{n_1}$ and $D_{n_2}$ functions define the vector $\bm{\xi}=\mathsf{G}\mathbf{r}$, with the matrix $\mathsf{G}=\sqrt{2}\mathsf{F}\mathsf{R}\mathsf{M}^{\nicefrac{1}{2}}$, and $\mathsf{F}=\diag{(\sqrt{\omega_1},\sqrt{\omega_2})}$. Explicitly, matrix $\mathsf{G}$ is given by
\begin{equation}\label{eq:G_matrix}
\mathsf{G}=\begin{pmatrix}
\sqrt{2m\omega_1}\cos{\theta} & \sqrt{2M\omega_1}\sin{\theta} \\ 
-\sqrt{2m\omega_2}\sin{\theta} & \sqrt{2M\omega_2}\cos{\theta}
\end{pmatrix}.
\end{equation}
The volume element in terms of the components of $\bm{\xi}$ is given by
\begin{equation}\label{eq:dxi_volume_element}
d\xi_1d\xi_2=2(1-\delta^2)^{\nicefrac{1}{4}}\sqrt{mM\omega\Omega}\,dxdy.
\end{equation}

\subsection{Born-Oppenheimer solution}\label{sec:BO_solution}

\subsubsection{The \texorpdfstring{$\chi_n(x,y)$}{} function}

The system $x$ has Hamiltonian $\hat{h}$
\begin{equation}\label{eq:Hamiltonian_x}
    \hat{h}=-\frac{1}{2m}\frac{\partial^2}{\partial x^2}+\frac{m\omega^2}{2} x^2+\frac{M\Omega^2}{2} y^2+ c x y.
\end{equation}
The $y$-parametrized TISE for $x$, $\hat{h}\chi=\varepsilon_n(y)\chi$, has analytical solution
\begin{equation}\label{eq:TISE_x_solution}
    \chi(x,y)=C(y)D_\nu(z)
\end{equation}
$C(y)$ is a constant function of $y$. $D_\nu(z)$ is a parabolic cylinder (Weber) 
function with
\begin{equation}\label{eq:z_definition}
    z(x,y)=\left(\sqrt{2m\omega}x+\sqrt{\frac{2M\Omega^2}{\omega}}\delta y\right),
\end{equation}
and
\begin{equation}\label{eq_nu_definition}
    \nu(y,\varepsilon)=\frac{\varepsilon}{\omega}-\frac{1}{2}+\frac{(\delta^2-1)M\Omega^2}{2\omega}y^2.
\end{equation}
The linear form of $z(x,y)$ allows to write
\begin{equation}\label{eq:z_definition_1}
    z(x,y)=(\partial_xz)x+(\partial_yz)y,
\end{equation}
with 
\begin{align}\label{eq:z_definition_2}
    \partial_xz&=\sqrt{2m\omega},\\
    \partial_yz&=\sqrt{\frac{2M\Omega^2}{\omega}}\delta,\\
    \partial_yx&=-\sqrt{\frac{m}{M}}\frac{\Omega\delta}{\omega}.
\end{align}

Bound solutions of \eqref{eq:Hamiltonian_x} are obtained by the condition $\nu(y,\varepsilon)=n$, with $n$ a non negative integer. It is easily shown that
\begin{equation}
    \varepsilon_n(y)=\omega\left(n+\frac{1}{2}\right)+\frac{(1-\delta^2)M\Omega^2}{2}y^2.
\end{equation}
These eigenvalues are the potential energy curves of the $y$ degree of freedom. Since $|\delta|<1$, the effect of the coupling can be seen as a reduction of $\Omega$ to an effective frequency $\Omega\sqrt{1-\delta^2}$. 

The inner product on $x$ of the $D_n(z)$ functions is obtained by using the chain rule,
\begin{equation}\label{eq:normalization_D}
\begin{split}
    \braket{D_l(z)|D_n(z)}&=\int_{-\infty}^{\infty}D_l(z)D_n(z)dx\\
    &=\int_{-\infty}^{\infty}D_lD_n(\partial_x z)^{-1}dz\\
    &=\frac{\sqrt{2\pi}}{\partial_x z}n!\delta_{ln}.
%    &=\sqrt{\frac{\pi}{m \omega}}n!\delta_{ln}.
\end{split}
\end{equation}
    
The normalized function $\chi_n$ is 
\begin{equation}
\begin{split}
    \chi_n(z)&=\frac{1}{\sqrt{n!}}\frac{\sqrt{\partial_x z}}{(2\pi)^{1/4}}D_n(z)\\
    &=\frac{1}{\sqrt{n!}}\left(\frac{m\omega}{\pi}\right)^{1/4}D_n(z).    
\end{split}
\end{equation}

\subsubsection{Phase-space equations of motion and stability analysis}

The equations of motion of $\left\{\varphi_n,\vartheta_n(y)\right\}$ are given by
\begin{equation}\label{eq:phase_space_varphi_n}
    \begin{split}
        \varphi_n'&=\vartheta_n,\\
    \vartheta_n'&=\left(-2M(\mathcal{E}-\varepsilon_n(y))-\beta_{nn}\right)\varphi_n,
    \end{split}
\end{equation}
with
\begin{align*}\label{eq:conditions_phase_space_varphi_n}
        \varepsilon_n(y)&=\omega\left(n+\tfrac{1}{2}+\tfrac{M\Omega^2}{2\omega}\left(1-\delta^2\right)y^2\right),\\
        \beta_{nn}&=-\frac{M\Omega^2\delta^2}{\omega}\left(n+\tfrac{1}{2}\right).
\end{align*}
A detailed calculation of the non-adiabatic coupling $\braket{\chi_n|\partial^2_y\chi_n}$ can be seen in the Appendix, section \ref{sec:appendix}. This system of two bilinearly coupled harmonic oscillators has a constant non-adiabatic coupling that only depends on the quantum number $n$. 

Equations \eqref{eq:phase_space_varphi_n} are the equations for the tangent field of $\{\varphi_n,\vartheta_n\}$. It is straightforward to verify that the fixed point of this field is $\{\varphi_n,\vartheta_n\}^*=\{0,0\}$. The eigenvalues of the equations for the tangent field are given by equation \eqref{eq:eigenvalues_A},
\begin{equation}\label{eq:eigenvalues_A_for HO}
    \lambda_{1,2}=\mp\sqrt{-2M(\mathcal{E}-\varepsilon_n(y))-\beta_{nn}}.
\end{equation}

The condition for elliptic behaviour of the phase-space trajectory is given for values of $y$ such that $\mathcal{E}>\varepsilon_n(y)-\frac{1}{2M}\beta_{nn}$, explicitly
\begin{equation}
        \mathcal{E}>\omega\left(n+\tfrac{1}{2}\right)\left(1+\frac{\Omega^2\delta^2}{2\omega^2}\right)+\frac{M\Omega^2}{2}\left(1-\delta^2\right)y^2.
\end{equation}
The effect of the non-adiabatic coupling is twofold. On one hand, the higher frequency oscillator displays an effective frequency $\tilde{\omega}_1=\omega\left(1+\frac{\Omega^2\delta^2}{2\omega^2}\right)$, with $\omega<\tilde{\omega}_1$, on the other hand, the lower frequency oscillator shows an effective frequency $\tilde{\omega}_2=\Omega\sqrt{1-\delta^2}$, with $\tilde{\omega}_2<\Omega$.

\subsubsection{The \texorpdfstring{$\varphi_{nl}(y)$}{} and \texorpdfstring{$\Psi_{nl}(x,y)$}{} functions}

Bound solutions of the system of equations \eqref{eq:phase_space_varphi_n} are given by $\varphi_{nl}=D_\mu(\tilde{y})$, with
\begin{align}\label{eq:mu_and_Bar_y}
    \mu&=-\frac{1}{2}\pm\frac{\mathcal{E}-\left(n+\frac{1}{2}\right)\left(\omega+\frac{\delta^2\Omega^2}{2\omega}\right)}{\Omega\sqrt{1-\delta^2}},\\
    \tilde{y}&=y\sqrt{2M\Omega}\left(1-\delta^2\right)^{1/4}.
\end{align}
The condition $\mu=l(n)$, with $l(n)=0,1,2,...$, gives the Born-Oppenheimer energy
\begin{equation}\label{eq:BO_energy_Bar_y}
\begin{split}
     \tilde{\mathcal{E}}_{nl}&=\omega\left(1+\frac{\delta^2\Omega^2}{2\omega^2}\right)\left(n+\tfrac{1}{2}\right)+\Omega\sqrt{1-\delta^2}\left(l+\tfrac{1}{2}\right)\\
     &=\tilde{\omega}_1\left(n+\tfrac{1}{2}\right)+\tilde{\omega}_2\left(l+\tfrac{1}{2}\right). 
\end{split}
\end{equation}
The second line of equation \eqref{eq:BO_energy_Bar_y} defines the Born-Oppenheimer frequencies, or energy differences between consecutive levels, $\tilde{\omega}_1$ and $\tilde{\omega}_2$. It can be proved that $\tilde\omega_2<\tilde\omega_1$ under the condition $\Omega<\omega$.  It is useful to define the matrix of BO square-root frequencies by $\tilde{\mathsf{F}}=\diag{(\sqrt{\tilde{\omega}_1},\sqrt{\tilde{\omega}_2})}$.

For the uncoupled case, a simple inspection of equation \eqref{eq:BO_energy_Bar_y} gives
\begin{equation}\label{eq:uncoupled_case_BO}
    \tilde{\mathcal{E}}_{nl}=\omega\left(n+\tfrac{1}{2}\right)+\Omega\left(l+\tfrac{1}{2}\right).
\end{equation}

The normalized solution for the $y$ degree of freedom is given by
\begin{equation}
\begin{split}\label{eq:Bar_y_varphi}
    \varphi_{nl}(\tilde{y})&=\frac{1}{\sqrt{l!}} \frac{\sqrt{\partial_y\tilde{y}}}{(2\pi)^{1/4}} D_{l}(\tilde{y})\\
    &=\frac{\left(4M^2\Omega^2\left(1-\delta^2\right)\right)^{1/8}}{\sqrt{l!}(2\pi)^{1/4}} D_{l}(\tilde{y}).
\end{split}
\end{equation}
The full Born-Oppenheimer product wavefunction is then
\begin{equation}
    \begin{split}\label{eq:full_BO_sol}
    \chi_n(z)\varphi_{nl}(\tilde{y})&=\frac{\left(4M^2\Omega^2\left(1-\delta^2\right)\right)^{1/8}}{\sqrt{l!}\sqrt{n!}} \left(\frac{m\omega}{2\pi^2}\right)^{1/4}D_n(z)D_{l}(\tilde{y})\\
    &=\frac{\left(mM\omega\Omega\sqrt{1-\delta^2}\right)^{1/4}}{\sqrt{\pi n!l!}}D_n(z)D_{l}(\tilde{y}).
\end{split}
\end{equation}

The  arguments of the parabolic cylinder functions $D_n$ and $D_l$, $z$ and $\tilde{y}$ respectively, define the vector, $\tilde{\bm{\xi}}=\begin{psmallmatrix} z \\ \tilde{y}\end{psmallmatrix}$, which is related to the position vector $\mathbf{r}$ by $\tilde{\bm{\xi}}=\tilde{\mathsf{G}}\mathbf{r}$ with 
\begin{equation}\label{eq:G_tilde_matrix}
    \tilde{\mathsf{G}}=\begin{pmatrix} \sqrt{2m\omega} & \sqrt{\frac{2M\Omega^2}{\omega}}\delta \\
    0 & \sqrt{2M\Omega}\left(1-\delta^2\right)^{\nicefrac{1}{4}} \end{pmatrix}.
\end{equation}
The volume element of the BO coordinates $\tilde{\bm{\xi}}$ is given by $dz d\tilde{y}=2(1-\delta^2)^{\nicefrac{1}{4}}\sqrt{m M \omega \Omega}\,dx dy$. The use of this volume element, and the product \eqref{eq:full_BO_sol}  gives the normalized BO solution

\begin{equation}
\label{eq:norm_BO_sol}
    \tilde{\Psi}_{nl}(z,\tilde{y})=\frac{1}{\sqrt{2\pi n!l!}}D_n(z)D_{l}(\tilde{y}).
\end{equation}

The vector of arguments of the exact and BO wavefunctions, $\bm{\xi}$ and $\tilde{\bm{\xi}}$ respectively, are related  by using the result $\mathbf{r}=\tilde{\mathsf{G}}^{-1}\tilde{\bm{\xi}}$ in the definition $\bm{\xi}=\mathsf{G}\mathbf{r}$, to obtain
\begin{equation}
    \bm{\xi}=\mathsf{G}\tilde{\mathsf{G}}^{-1}\tilde{\bm{\xi}}.
\end{equation}
It can be proved that the determinant of the product $\tilde{\mathsf{G}}^{-1}$ is one. In analogy to the exact solution \eqref{eq:Psi_analytical_wf}, the matrix $\tilde{\mathsf{G}}$ can be written as the product 
$\tilde{\mathsf{G}}=\sqrt{2}\tilde{\mathsf{F}}\tilde{\mathsf{R}}\mathsf{M}^{\nicefrac{1}{2}}$, which allows to obtain $\tilde{\mathsf{R}}$ as
\begin{align}\label{R tilde equation}
    \tilde{\mathsf{R}}&=\tfrac{1}{\sqrt{2}}\tilde{\mathsf{F}}^{-1}\tilde{\mathsf{G}}\mathsf{M}^{\nicefrac{-1}{2}}\\
    &=\begin{pmatrix} \frac{\omega}{\sqrt{\omega^2+\delta^2\Omega^2/2}} & \frac{\delta\Omega}{\sqrt{\omega^2+\delta^2\Omega^2/2}} \\
    0 & 1 \end{pmatrix}.
\end{align} 
This upper triangular matrix is clearly non orthogonal as matrix $\mathsf{R}$.

\subsection{Error analysis of the Born-Oppenheimer solutions}

\subsubsection{Error of the Born-Oppenheimer energies}

The exact and BO energies for the uncoupled case, equations \eqref{eq:analytical_energy_uncoupled} and \eqref{eq:uncoupled_case_BO} respectively, are identical if the replacements $n_1\to n$ and $n_2 \to l$ is made on equation \eqref{eq:analytical_energy_uncoupled}. In the general case, equation \eqref{eq:analytical_energy}, the exact energies are given by
\begin{equation}\label{eq:analytical_energy_nl}
    \mathcal{E}_{nl}=\omega_1\left(n+\tfrac{1}{2}\right)+\omega_2\left(l+\tfrac{1}{2}\right).
\end{equation}

\begin{figure}[htb]
    \centering
    \includegraphics[scale=0.75]{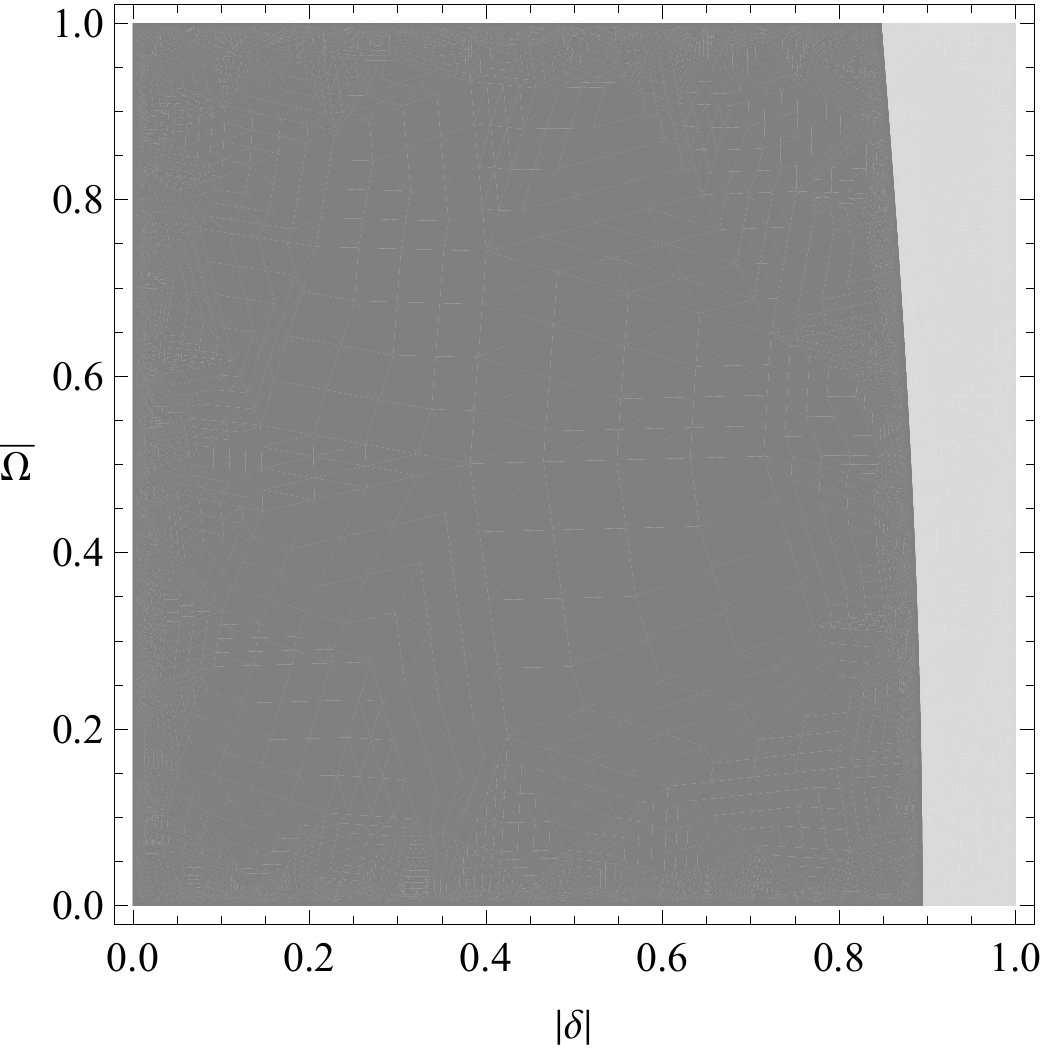}
    \caption{Regions of the $\delta-\bar{\Omega}$ plane where the Born-Oppenheimer approximation overestimates, $\epsilon_\omega>0$ in light gray, or underestimate, $\epsilon_\omega<0$ in dark gray, the exact energy $\omega_2$. The BO energies overestimate the exact energy of the $\omega_2$ oscillator at large values of the coupling $\delta$. This figure is used also to show the regions of the $\delta-\bar{\Omega}$ plane where the sign of the slope of line \eqref{eq:zero_epsilon_line} is positive (dark gray) or negative (light gray).}
    \label{fig:tilde_epsilon_omega}
\end{figure}

The relative error in the BO energy levels is given by
\begin{equation}\label{eq:BO_error}
\begin{split}
    \epsilon_{nl}^{BO}&=\frac{\mathcal{E}_{nl}^{BO}-\mathcal{E}_{nl}}{\mathcal{E}_{nl}}\\
    &=\frac{\epsilon_\omega\left(n+\tfrac{1}{2}\right)+\epsilon_\Omega\left(l+\tfrac{1}{2}\right)}{\bar{\omega}_1\left(n+\tfrac{1}{2}\right)+\bar{\omega}_2\left(l+\tfrac{1}{2}\right)}.
    \end{split}
\end{equation}
with the definitions
\begin{align}\label{eq:reduced_epsilon_omega}
    \epsilon_\omega&=\omega^{-1}\left(\tilde{\omega}_1-\omega_1\right)\\
    &=\left(1+\tfrac{\delta^2\bar{\Omega}^2}{2}\right)-\bar{\omega}_1,\\
    \epsilon_\Omega&=\omega^{-1}\left(\tilde\omega_2-\omega_2\right)\\
    &=\bar{\Omega}\sqrt{1-\delta^2}-\bar{\omega}_2,
\end{align}
and 
\begin{equation}\label{eq:reduced_omega_12}
\begin{split}
    \bar{\omega}_{1,2}&=\omega_{1,2}/\omega\\
    &=\tfrac{1}{\sqrt{2}}\left(1+\bar{\Omega}^2\pm\sqrt{\left(1-\bar{\Omega}^2\right)^2+4\delta^2\bar{\Omega}^2}\right)^{1/2},  
    \end{split}
\end{equation}
in terms of $\bar{\Omega}=\Omega/\omega$. The relative error of the BO energies, equation \eqref{eq:BO_error}, defines a surface in terms of $n$ and $l$, for a given value of $\delta$ and $\bar{\Omega}$. The magnitude and sign of the BO error can be analyzed by a detailed examination of the terms of equation \eqref{eq:BO_error}. Since the denominator of $\epsilon_{nl}^{BO}$ is a positive quantity, the analysis of the sign of the BO error depends entirely on the two terms of the numerator of \eqref{eq:BO_error}. For the second term of the numerator, the inequality $\epsilon_\Omega>0$ leads to the result $\delta^2\Omega^2(\delta^2-1)<0$, which holds for $0<|\delta|<1$ , indicating that the BO result overestimates the energy of the $\omega_2$ oscillator for any value of the parameters $\delta$ and $\bar{\Omega}$. For the first term of the numerator, the quantity $\epsilon_\omega$ is greater, or less, than zero in the region of the $\delta-\bar{\Omega}$ plane implicitly defined by $\mathcal{B}(\delta,\bar{\Omega})>0$, or $\mathcal{B}(\delta,\bar{\Omega})<0$ respectively, with $\mathcal{B}(\delta,\bar{\Omega})=4(5\delta^2-4)+4(2\delta^2-1)\bar{\Omega}^2+\delta^6\bar{\Omega}^4$, and $\bar{\Omega}<1$. Figure \ref{fig:tilde_epsilon_omega} displays regions of the $\delta-\bar{\Omega}$ plane where the BO energy overestimates $\epsilon_\omega>0$, or underestimates $\epsilon_\omega<0$, the energy of the oscillator $\omega_1$, it can be seen in this figure, that the BO energy overestimate the exact energy $\omega_1$ for large values of the coupling $\delta$.

\begin{figure}[htb]
    \centering
    \includegraphics[scale=0.75]{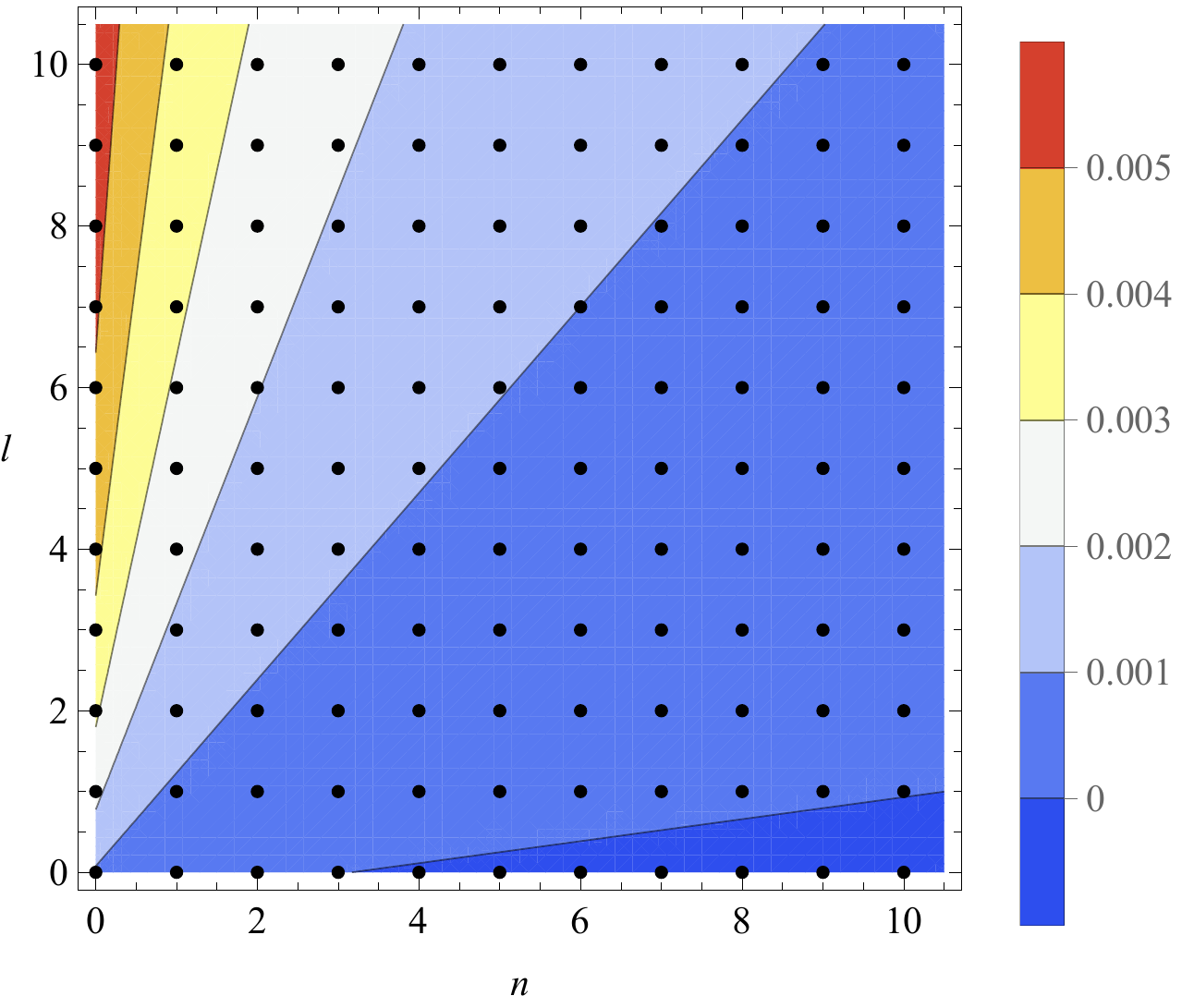}
    \caption{Contours of constant relative error, equation \eqref{eq:BO_error}, in the $n-l$ plane for $\delta=0.6$ and $\bar{\Omega}=0.2$. The lattice of quantum states is displayed as black points. Colored regions, as indicated by the legend, give the range of the error for each quantum state.}
    \label{fig:relative_error_nl}
\end{figure}

The boundary between regions of the $n-l$ plane with positive and negative relative error is given by the condition $\epsilon^{BO}=0$. This condition gives the equation of a straight line, 
\begin{equation}\label{eq:zero_epsilon_line}
    l=-\frac{\epsilon_\omega}{\epsilon_\Omega}n-\frac{1}{2}\left(1+\frac{\epsilon_\omega}{\epsilon_\Omega}\right),
\end{equation}
with slope $-\epsilon_\omega/\epsilon_\Omega$, and $l$-intercept $-\frac{1}{2}\left(1+\frac{\epsilon_\omega}{\epsilon_\Omega}\right)$. Since $\epsilon_\Omega>0$, the sign of the slope of line \eqref{eq:zero_epsilon_line} depends only on $\sgn(-{\epsilon_\omega})$. Figure \ref{fig:tilde_epsilon_omega} displays the regions of the $\delta-\bar{\Omega}$ plane where the error $\epsilon_\omega$ is positive or negative, this figure also shows the regions of the $\delta-\bar{\Omega}$ plane where the slope of the line \eqref{eq:zero_epsilon_line} is positive or negative. It can be seen in Figure \ref{fig:tilde_epsilon_omega} that the slope of line \eqref{eq:zero_epsilon_line} is positive only for large values of $\delta$. It can be proved after some algebraic manipulations that the inequality $\left(1+\frac{\epsilon_\omega}{\epsilon_\Omega}\right)>0$
holds for $0<\bar{\Omega}<1$ and $0<|\delta|<1$, which shows that the upper limit of the slope is one, i.e. $-\frac{\epsilon_\omega}{\epsilon_\Omega}<1$, and that the $l$-intercept is always negative. These two results about the slope and the $l$-intercept of line \eqref{eq:zero_epsilon_line} show that the BO approximation overestimates the energy of the ground state for linearly coupled oscillators. The overestimation of the BO ground state energy contrasts with previous works: Bratsev \cite{Bratsev1965}, Epstein \cite{Epstein1966} and McCoy \cite{McCoy2001}, who obtained that the BO ground state energy is always smaller than the true ground state energy. Figure \ref{fig:relative_error_nl} displays the contours of constant relative error $\epsilon_{nl}^{BO}$ as functions of $n$ and $l$ for $\delta=0.6$ and $\bar{\Omega}=0.2$, the zero contour, $\epsilon_{nl}^{BO}=0$, gives the straight line \eqref{eq:zero_epsilon_line}. The relative error of the BO states are less than one percent, as can be seen in figure \ref{fig:relative_error_nl} with the aid of the legend. In figure \ref{fig:relative_error_nl} it is seen that BO states with $l=0$ display the smallest relative error, while the quantum states with $n=0$ show the biggest relative errors.

\subsubsection{Wave functions: error analysis}

\begin{figure}[htb]
    \centering
    \includegraphics[scale=0.75]{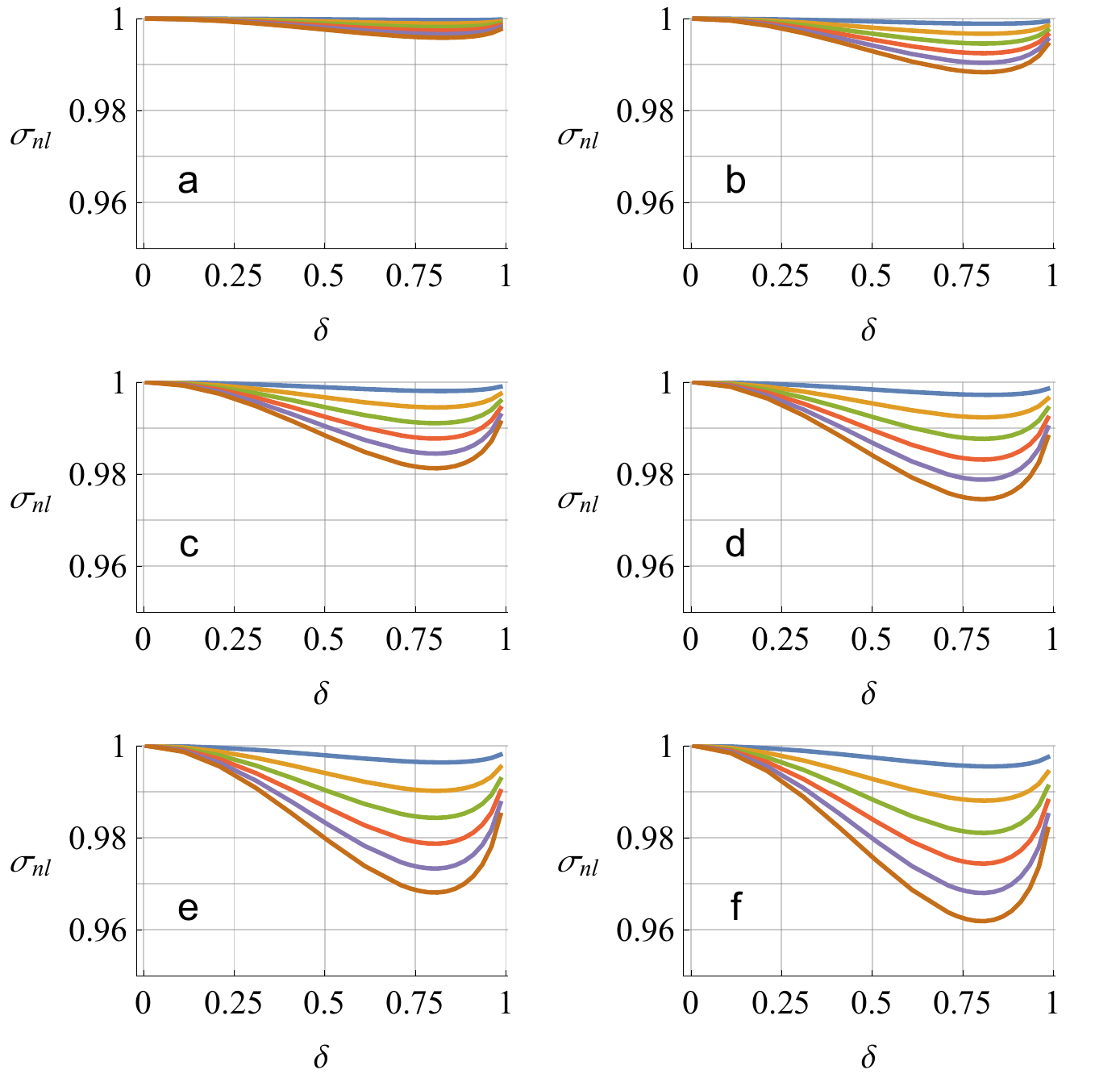}
    \caption{The inner product $\sigma_{nl}$ as a function of $\delta$ for $\bar{\Omega}=0.2$. In all the panels: the blue, orange, green, red, purple, and brown curves represent the cases $l=0,1,2,3,4,5$, respectively. Panels a, b, c, d, e, and f, correspond to $n=0,1,2,3,4$, and $5$, respectively. The smallest error is given by the ground state, $\sigma_{00}$. For a given value of $\delta$ the $\sigma_{nl}$ decreases as the quantum number $n$ and $l$ increase. }
    \label{fig:wf_overlap}
\end{figure}

The exact and BO wave functions, equations \eqref{eq:Psi_analytical_wf} and \eqref{eq:full_BO_sol}, are products of parabolic cylinder functions $D_\nu$, with $\nu$ non negative integer. Since the $D_\nu$ are $L^2$ functions, the exact and the BO functions must be $L^2$ functions, allowing to use the inner product $\sigma_{nl}=\braket{\Psi_{nl}|\tilde{\Psi}_{nl}}$ as a measure of error of the BO function compared to the exact wavefunctions. Figure \ref{fig:wf_overlap} displays the value of $\sigma_{nl}$ as function of $\delta$ for $\bar{\Omega}=0.2$. In all the panels of the figure, the value of $\sigma_{nl}$ is very close to one, which reinforces the observed in figure \ref{fig:relative_error_nl}, and shows that the BO approximation works very well for linearly coupled harmonic oscillators. In all the panels of this figure is clear that for a given value of $\delta$, the higher the values of $n$ and $l$, the more the BO and exact wavefunctions differ. It is interesting to compare the relative error of figure \ref{fig:relative_error_nl} with the wavefunction overlap of figure \ref{fig:wf_overlap} for the same values of $n$ and $l$. The relative error for $n=0$ displays the largest values, meanwhile the wavefunction overlap displays larger values as $n$ and $l$ grow larger.

\begin{figure}[htb]
    \centering
    \includegraphics[scale=0.75]{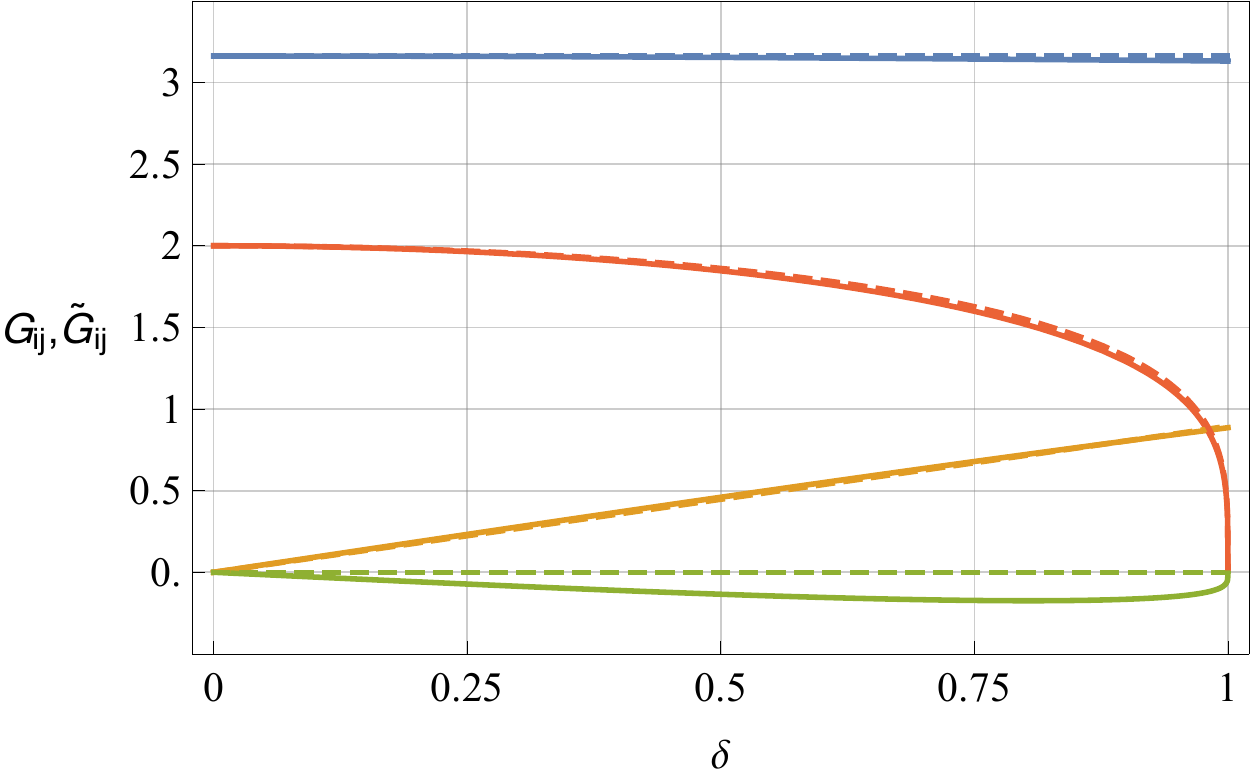}
    \caption{Components of the matrices $\mathsf{G}$ (solid) and $\tilde{\mathsf{G}}$ (dashed) as functions of the coupling parameter $\delta$ for $\bar{\Omega}=0.2$. Components $\mathsf{G}_{11}$, $\mathsf{G}_{12}$, $\mathsf{G}_{21}$ and $\mathsf{G}_{22}$ are displayed in blue, orange, green and red, respectively, with same color code for the elements of $\tilde{\mathsf{G}}$.}
    \label{fig:G_barG_components}
\end{figure}

All the panels of figure \ref{fig:wf_overlap} show that $\lim_{\delta\to 0}\sigma_{nl}=1$, and $\lim_{\delta\to 1}\sigma_{nl}=1$. This can be explained by the vectors of arguments of the parabolic cylinder functions: $\bm{\xi}=\mathsf{G}\mathbf{r}$, and $\tilde{\bm{\xi}}=\tilde{\mathsf{G}}\mathbf{r}$. The components of the transformation matrices $\mathsf{G}$ and $\tilde{\mathsf{G}}$ are displayed in figure \ref{fig:G_barG_components}. Three of the four elements of the BO transformation matrix $\tilde{\mathsf{G}}$ display excellent agreement with the elements of $\mathsf{G}$. Equation \eqref{eq:G_tilde_matrix} show that the BO $\tilde{\mathsf{G}}_{21}$ element is always zero, as shown by the dashed green curve of figure \ref{fig:G_barG_components}. The green solid curve shows that $\mathsf{G}_{21}$ is negative for all values of  $\delta$, being zero only at the endpoints $\delta=0$ and $\delta=1$. The component $\mathsf{G}_{21}$ displays
its lowest value for $\delta\approx0.8$, which agrees fairly well with the panels of figure \ref{fig:wf_overlap}. More precisely, the matrix element $\mathsf{G}_{21}$ can be minimized with respect to $\delta$ to obtain
\begin{equation}\label{eq:delta_star}
    \delta^*=\frac{\sqrt{\left(1-\bar{\Omega}^2\right)\left(3+5\bar{\Omega}^2-\sqrt{9-2\bar{\Omega}^2-7\bar{\Omega}^4}\right)}}{2\sqrt{2}\bar{\Omega}},
\end{equation}
with $\bar{\Omega}=\Omega/\omega$. The case of figure \ref{fig:G_barG_components}, with $\bar{\Omega}=0.2$, gives $\delta^*=\tfrac{2}{5}\sqrt{60-6\sqrt{87}}\approx0.803565$, and $\mathsf{G}^*_{21}=-0.174359\sqrt{m}$. 

\subsection{\texorpdfstring{$\mathsf{QU}$}{} decomposition of matrix \texorpdfstring{$\mathsf{G}$}{}}

\begin{figure}[htb]
    \centering
    \includegraphics[scale=0.75]{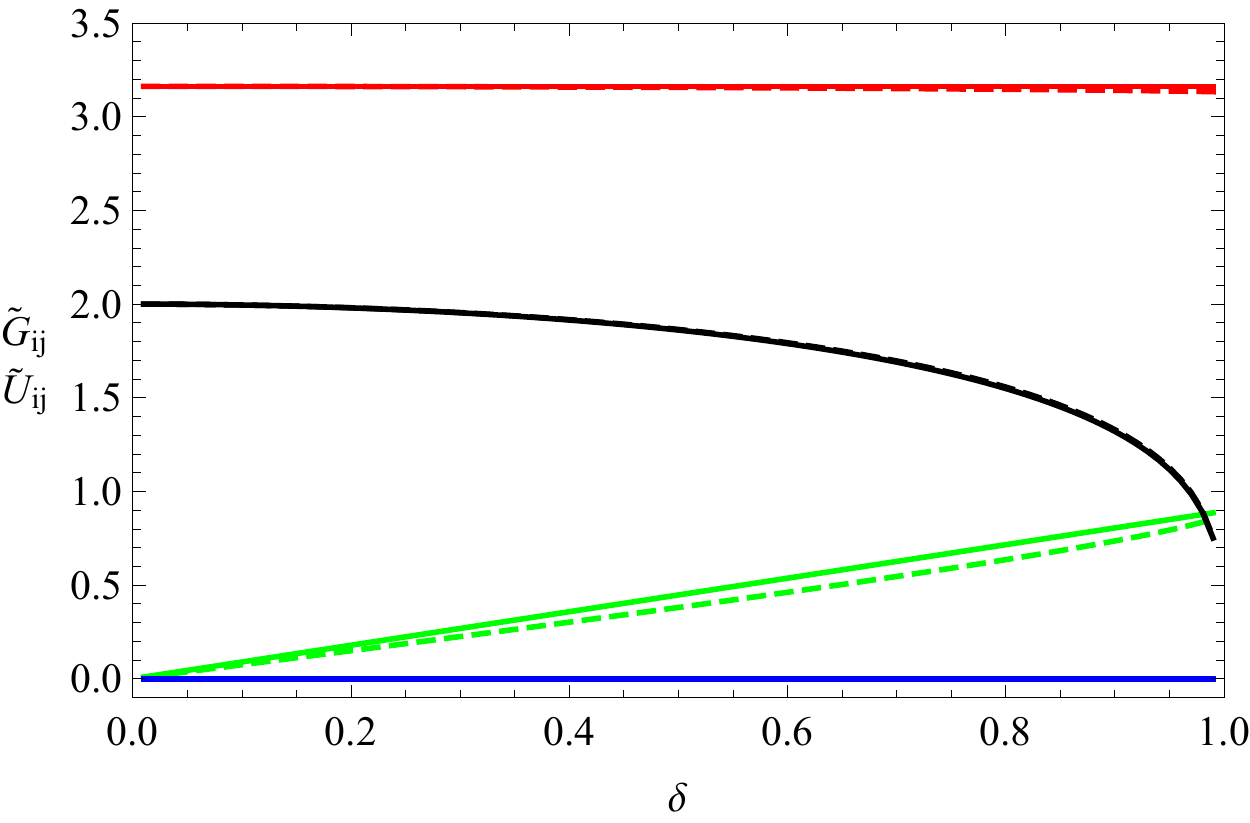}
    \caption{Elements of matrices $\mathsf{\tilde{G}}$ and $\mathsf{\tilde{U}}$ as function of $\delta$ for $\bar{\Omega}=0.2$. Solid and dashed curves represent $\mathsf{\tilde{G}}$ and $\mathsf{\tilde{U}}$ respectively. $\tilde{G}_{11}$ and $\tilde{U}_{11}$ are displayed in red, $\tilde{G}_{12}$ and $\tilde{U}_{12}$ in green,
    $\tilde{G}_{21}$ and $\tilde{U}_{21}$ in blue, and $\tilde{G}_{22}$ and $\tilde{U}_{22}$ in black. }
    \label{fig:QU_analysis}
\end{figure}

Matrices $\mathsf{G}$ and $\tilde{\mathsf{G}}$, equations \eqref{eq:G_matrix} and \eqref{eq:G_tilde_matrix} respectively, are related by the $\mathsf{QU}$ decomposition of $\mathsf{G}$ \cite{Strang2019},
\begin{equation}\label{eq:QU_decomposition}
    \mathsf{G}=\mathsf{QU},
\end{equation}
with $\mathsf{Q}$ an orthogonal matrix and $\mathsf{U}$ an upper triangular matrix. Let define the matrix   $\mathsf{\tilde{U}}=-\sigma_3\mathsf{U}$, with $\sigma_3=\diag{(1,-1)}$ as the Pauli's diagonal matrix. It is observed that the BO matrix $\tilde{\mathsf{G}}$ is approximately given by $\mathsf{\tilde{U}}$, $\tilde{\mathsf{G}}\approx\mathsf{\tilde{U}}$. Figure \ref{fig:QU_analysis} displays the elements of matrices $\mathsf{\tilde{G}}$ and $\mathsf{\tilde{U}}$, $\tilde{G}_{ij}$ and $\tilde{U}_{ij}$ respectively, as function of $\delta$ for $\bar{\Omega} = 0.2$. As the value of $\bar{\Omega}$  increases, the approximation $\tilde{\mathsf{G}}\approx\mathsf{\tilde{U}}$ becomes more accurate. 

The $\mathsf{QU}$ decomposition can be used to improve the BO wavefunction. The fact that $\tilde{\mathsf{G}}\approx\mathsf{\tilde{U}}$ can be used to improve the BO matrix $\tilde{\mathsf{G}}$ by using the $\mathsf{QU}$ decomposition inversely. Multiplying the matrix $\mathsf{U}=-\sigma_3\tilde{\mathsf{G}}$ by an orthogonal transformation $\mathsf{Q}$, for example the rotation matrix $\mathsf{Q}(\phi)$, is possible to improve the Born-Oppenheimer approximation. This approach is being analyzed in further details by the author and collaborators.

\section{Conclusions}

It has been proposed a method for finding bound states, for systems with two degrees of freedom, based on the BO factorization ansatz and the phase-space bound trajectory approach. This method has been tested by using a system consisting of two 1-dimensional harmonic oscillators bilinearly coupled. The use of bilinearly coupled harmonic oscillators makes straightforward to solve the phase-space differential equations for both oscillators, since the non-adiabatic coupling $\beta_{nn}$ is a constant. The systems of bilinearly coupled oscillators is analytically solvable in terms of a linear transformation of coordinates, this is also the case for the solution obtained by using the Born-Oppenheimer ansatz and the phase-space bound trajectory method. The analytical non-adiabatic eigenvalues and eigenfunctions have been compared with the analytical solutions showing an excellent accuracy of the BO-phase-space trajectory approach for the system of two bilinearly coupled harmonic oscillators. An extensive mathematical analysis of the energy eigenvalues and their relative errors has been done. This analysis has been useful identifying regions on the parameter space where the BO approximation overestimate or underestimate the exact quantum energy. The relative error of the BO approximation for the lattice of quantum states has been analyzed in detail. It has been shown that the relative error is mostly positive and grows larger as the quantum numbers increase. The Born-Oppenheimer wave functions are compared with the exact quantum wavefunctions by using an overlap integral as a measure of error. The error of the BO wavefunctions has been obtained as a function of the coupling parameters $\delta$. The use of transformation matrices between the physical coordinates and the arguments of the parabolic cylinder functions has allowed to analyze in detail the error in the wavefunction by comparing the quantum and BO matrix elements, matrices $\mathsf{G}$ and $\mathsf{\tilde{G}}$ respectively. A $\mathsf{QU}$ decomposition of matrix $\mathsf{G}$ outputs a upper triangular matrix $\mathsf{U}$ which is closely related to the BO transformation matrix $\mathsf{\tilde{G}}$. The use of the $\mathsf{QU}$ decomposition to improve the Born-Oppenheimer approximation has been outlined, this issue is currently investigated by the author and collaborators.

\section{Acknowledgments}
This project has been fully financed by the internal research grants of Universidad Icesi. 

\section{Data Availability Statement}

The data that support the findings of this study are available from the corresponding author upon reasonable request.

\section{Appendix: Calculations with the \texorpdfstring{$D_\nu$}{} functions}\label{sec:appendix}

This appendix displays some of the mathematical facts, equations, and procedures used in section \ref{sec:BO_solution}

\subsection{Recurrence relations for the \texorpdfstring{$D_\nu$}{} functions}  

\begin{align}\label{eq:recurrence_D}
    D_{\nu+1}(z)-z D_{\nu}(z)+\nu D_{\nu-1}(z)&=0,\\
    D'_{\nu}(z)+\tfrac{z}{2}D_\nu(z)-\nu D_{\nu-1}(z)&=0,\\
    D'_{\nu}(z)-\tfrac{z}{2}D_\nu(z)+D_{\nu+1}(z)&=0.        
\end{align}

\subsection{Matrix elements of the \texorpdfstring{$D_\nu$}{} functions}

The matrix elements $\braket{D_l(z)|xD_n(z)}$ and $\braket{D_l(z)|x^2D_n(z)}$ (integral over $x$) are obtained from the recurrence relations \eqref{eq:recurrence_D}, and the definition of $z$ given by equation \eqref{eq:z_definition} and \eqref{eq:z_definition_2},
\begin{align}
    \braket{D_l(z)|xD_n(z)}&=\frac{1}{\partial_xz}\int_{-\infty}^\infty{xD_lD_ndz}\\
    &=\frac{1}{(\partial_xz)^2}\int_{-\infty}^\infty{\left(z-(\partial_yz)y\right)D_lD_ndz}\\
    &=\frac{1}{(\partial_xz)^2}\int_{-\infty}^\infty{zD_l D_n dz}-\frac{(\partial_yz)y}{(\partial_xz)^2}\int_{-\infty}^\infty{D_lD_ndz}\\
    &=\frac{(\partial_yx)y}{(\partial_xz)}\sqrt{2\pi}n!\delta_{ln},
\end{align}
and
\begin{align}
    \braket{D_l(z)|x^2 D_n(z)}&=\frac{1}{\partial_xz}\int_{-\infty}^\infty{x^2 D_l D_ndz}\\
    &=\frac{1}{(\partial_xz)^2}\int_{-\infty}^\infty{\left(z-(\partial_yz)y\right)^2 D_l D_ndz}\\
    &=\frac{1}{(\partial_xz)^2}\int_{-\infty}^\infty{z^2D_l D_ndz}+\frac{(\partial_yz)^2y^2}{(\partial_xz)^2}\int_{-\infty}^\infty{D_lD_ndz}\\
    &=\left(\frac{y\partial_yz}{\partial_xz}\right)^2\sqrt{2\pi}n!\delta_{ln}+\frac{1}{(\partial_xz)^2}\int_{-\infty}^\infty{(z D_l)(zD_n)dz}\\
    &=(y\partial_yx)^2\sqrt{2\pi}n!\delta_{ln}+\frac{1}{(\partial_xz)^2}\int_{-\infty}^\infty{(D_{l+1}+lD_{l-1})(D_{n+1}+nD_{n-1})dz}\\
    &=(y\partial_yx)^2\sqrt{2\pi}n!\delta_{ln}\\
    &+\frac{1}{(\partial_xz)^2}\int_{-\infty}^\infty{(D_{l+1}D_{n+1}+nD_{l+1}D_{n-1}+lD_{l-1}D_{n+1}+lnD_{l-1}D_{n-1})dz}\\  
    &=(y\partial_yx)^2\sqrt{2\pi}n!\delta_{ln}\\
    &+\frac{\sqrt{2\pi}}{(\partial_xz)^2}\left((n+1)!\delta_{l+1,n+1}+n!\delta_{l+1,n-1}+l(n+1)!\delta_{l-1,n+1}+ln!\delta_{l-1,n-1}\right)\\
    &=(y\partial_yx)^2\sqrt{2\pi}n!\delta_{ln}\\
    &+\frac{\sqrt{2\pi}n!}{(\partial_xz)^2}\left((n+1)\delta_{l+1,n+1}+\delta_{l+1,n-1}+l(n+1)\delta_{l-1,n+1}+l\delta_{l-1,n-1}\right)
\end{align}

\subsection{Derivatives of the \texorpdfstring{$\chi(z)$}{} function}

By using the chain rule, and the recurrence relations \eqref{eq:recurrence_D}, we obtain the derivatives of $\chi_n(z)$,
\begin{equation}\label{eq:first_derivative_D_1}
\begin{split}
    \partial_y\chi_n(z)&=\frac{1}{\sqrt{n!}}\frac{\sqrt{\partial_x z}}{(2\pi)^{1/4}}D_n'(\partial_yz)\\
    &=\frac{(\partial_yz)}{\sqrt{n!}}\frac{\sqrt{\partial_x z}}{(2\pi)^{1/4}}\left(\tfrac{1}{2}zD_n-D_{n+1}\right)\\
    &=\frac{(\partial_yz)}{2\sqrt{n!}}\frac{\sqrt{\partial_x z}}{(2\pi)^{1/4}}\left(n D_{n-1}-D_{n+1}\right)\\
    &=\frac{(\partial_yz)}{2}\left(\sqrt{n}\chi_{n-1}-\sqrt{n+1}\chi_{n+1}\right)
    %\sqrt{\frac{2 c^2}{m\omega^3}}
\end{split}
\end{equation}
and
\begin{equation}\label{eq:second_derivative_D}
\begin{split}
    \partial^2_y\chi_n(z)&=\frac{1}{\sqrt{n!}}\frac{\sqrt{\partial_x z}}{(2\pi)^{1/4}}\partial_y(D'_n\partial_yz)\\
    &=\frac{1}{\sqrt{n!}}\frac{\sqrt{\partial_x z}}{(2\pi)^{1/4}}(D''_n(\partial_yz)^2+D'_n\partial_y^2z)\\
    &=\frac{(\partial_yz)^2}{\sqrt{n!}}\frac{\sqrt{\partial_x z}}{(2\pi)^{1/4}}D''_n\\
    &=-\frac{\left(\partial_yz\right)^2}{\sqrt{n!}}\frac{\sqrt{\partial_x z}}{(2\pi)^{1/4}}\left((n+\tfrac{1}{2})-(\tfrac{z}{2})^2\right)D_n\\
    &=-\left((n+\tfrac{1}{2})-(\tfrac{z}{2})^2\right)\left(\partial_yz\right)^2\chi_n.
\end{split}
\end{equation}

\subsubsection{Matrix elements of \texorpdfstring{$\chi_n(x,y)$}{} }

The use of the chain rule, and the recurrence relations \eqref{eq:recurrence_D}, allow to obtain the derivatives of $\chi_n(z)$,
\begin{equation}\label{eq:first_derivative_D}
\begin{split}
    \partial_y\chi_n(z)&=\frac{(\partial_yz)}{2}\left(\sqrt{n}\chi_{n-1}-\sqrt{n+1}\chi_{n+1}\right)\\
    \partial^2_y\chi_n(z)&=-\left((n+\tfrac{1}{2})-(\tfrac{z}{2})^2\right)\left(\partial_yz\right)^2\chi_n.
    %\sqrt{\frac{2 c^2}{m\omega^3}}
\end{split}
\end{equation}

The matrix elements of equations \eqref{eq:vartheta_equation_diagonal} and \eqref{eq:vartheta_equation_off_diagonal} are 
\begin{equation}\label{eq:matrix_element_1st_derivative}
\begin{split}
    \braket{\chi_l|\partial_y\chi_n}&=\frac{(\partial_yz)}{2}\left(\sqrt{n}\braket{\chi_l|\chi_{n-1}}-\sqrt{n+1}\braket{\chi_l|\chi_{n+1}}\right)\\
    &=\frac{(\partial_yz)}{2}\left(\sqrt{n}\delta_{l,n-1}-\sqrt{n+1}\delta_{l,n+1}\right),
\end{split}
\end{equation}
and
\begin{equation}\label{eq:matrix_element_2nd_derivative}
\begin{split}
    \braket{\chi_l|\partial^2_y\chi_n}&=-\left(\partial_yz\right)^2\left((n+\tfrac{1}{2})\braket{\chi_l|\chi_n}-\tfrac{1}{4}\braket{\chi_l|z^2\chi_n}\right)\\
    &=-\left(\partial_yz\right)^2\left((n+\tfrac{1}{2})\delta_{l,n}-\tfrac{1}{4}\braket{\chi_l|z^2\chi_n}\right).
\end{split}
\end{equation}
Explicitly, the last integral on the RHS of equation \eqref{eq:matrix_element_2nd_derivative} is given by
\begin{equation}\label{eq:matrix_element_chi_2nd_derivative}
\begin{split}
     \braket{\chi_l|z^2\chi_n}&=\frac{(\partial_x z)}{\sqrt{2\pi l!n!}}\int_{-\infty}^{\infty}{D_l(z)z^2D_n(z)dx}\\
     &=\frac{1}{\sqrt{2\pi l!n!}}\int_{-\infty}^{\infty}{z^2 D_l D_ndz}\\
     &=\frac{\sqrt{n!}}{\sqrt{l!}}\left((n+1)\delta_{l+1,n+1}+\delta_{l+1,n-1}+l(n+1)\delta_{l-1,n+1}+l\delta_{l-1,n-1}\right).
\end{split}
\end{equation}
The diagonal elements of matrices \eqref{eq:matrix_element_1st_derivative} and \eqref{eq:matrix_element_2nd_derivative} are $\braket{\chi_n|\partial_y\chi_n}=0$, and 
\begin{equation}\label{eq:diagonal_matrix_chi_1st_derivative}
    \braket{\chi_n|\partial^2_y\chi_n}=-\left(\frac{\partial_yz}{2}\right)^2(2n+1).  
\end{equation}
The nonzero off-diagonal elements of matrices \eqref{eq:matrix_element_1st_derivative} and \eqref{eq:matrix_element_2nd_derivative} are
\begin{align}\label{eq:off_diagonal_matrix_chi_1st_derivative}
    \braket{\chi_{n-1}|\partial_y\chi_n}&=\frac{\partial_yz}{2}\sqrt{n},\\
    \braket{\chi_{n+1}|\partial_y\chi_n}&=-\frac{\partial_yz}{2}\sqrt{n+1},
\end{align}
and
\begin{align}\label{eq:off_diagonal_matrix_chi_2nd_derivative}
    \braket{\chi_{n-2}|\partial^2_y\chi_n}&=\left(\frac{\partial_yz}{2}\right)^2n(n-1),\\
    \braket{\chi_{n+2}|\partial^2_y\chi_n}&=\left(\frac{\partial_yz}{2}\right)^2.
\end{align}
The use of these matrix elements in equation \eqref{eq:vartheta_equation_off_diagonal} produces
\begin{align}\label{eq:offdiagonal_matrix_equation_chi_1}
    -\braket{\chi_{n-1}|\partial_y^2\chi_n}\varphi_n&=2\braket{\chi_{n-1}|\partial_y\chi_n}\vartheta_n\\
    0&=(\partial_yz)\sqrt{n}\vartheta_n\\
    \vartheta_n&=0.
\end{align}
and 
\begin{align}\label{eq:offdiagonal_matrix_equation_chi_2}
    -\braket{\chi_{n+1}|\partial_y^2\chi_n}\varphi_n&=2\braket{\chi_{n+1}|\partial_y\chi_n}\vartheta_n\\
    0&=-(\partial_yz)\sqrt{n+1}\vartheta_n\\
    \vartheta_n&=0.
\end{align}

\bibliographystyle{aipnum4-1}
\bibliography{main}
\end{document}